\documentclass[report,floatfix,aps,nofootinbib]{revtex4}
\usepackage{epsfig}

\def\lp {\left( }
\def\rp {\right) }
\def\lb {\left[ }
\def\rb {\right] }

\def\nn {\nonumber}

\def\beq{\begin{equation}}
\def\eeq{\end{equation}}
\def\bea{\begin{eqnarray}}
\def\eea{\end{eqnarray}}
\def\ni{\noindent}

\def\m{\mu}
\def\n{\nu}

\begin{document}

\title{Quantum mechanics in curved space-time}

\author{C.C. Barros Jr.}

\affiliation{
Instituto de F\'{\i}sica, Universidade de S\~{a}o Paulo,\\
C.P. 66318, 05315-970, S\~ao Paulo, SP, Brazil}


\begin{abstract}
In this paper, the principles of the general 
relativity are used to formulate quantum wave equations 
for spin-0 and spin-1/2 particles. More specifically, the equations are worked
in a Schwarzschild-like metric.
As a test, the hydrogen atom spectrum is calculated. 
A comparison of the calculated spectrum with the numerical data of the
 deuterium energy levels
shows a significant improvement of the accord, and the deviations 
are almost five times smaller then the ones obtained with the Dirac theory. 
The implications of the 
theory considering the strong interactions are also discussed.
\end{abstract}

\maketitle

\vspace{5mm}

\section{introduction}

The  general theory of the relativity, proposed in 1916  by Einstein, was one 
of the 
major scientific discoveries of last century. Besides providing very accurate
theoretical results, it was a great advance in the understanding the Nature,
dealing with the structure of the space-time.

A question of interest is how the quantum theory can be affected by the
space-time.
Dirac, formulated his theory \cite{dir1}, based in the flat 
space-time of the special theory of the relativity, and with this formulation,
the spin and the antiparticles appeared naturally into the theory. 

About curved spaces, many authors, as for example \cite{rfd}, \cite{dwit}, 
\cite{hawk1} proposed methods to quantize the gravity, where still there
are many difficulties and opened questions to be understood.
 
In this work, a different point of view is proposed. Here, instead of trying 
to quantize
 the gravity, 
the effects of the metric in the subatomic world will be studied.
For this purpose, the
basic idea is to describe a particle in a region with  a potential that
affects the metric of the space-time.
We are not interested in gravitational effects, as in \cite{parker}, 
where the effect of gravitational forces in the hydrogen atom spectrum has
been included. So, the gravitational potential will be turned off
and only the other interactions (strong, electromagnetic) will be considered. 
Observing that  the masses of the particles are very small, and the
small value of the gravitational coupling, when compared with
the electric or strong ones, one can say that it is an excellent 
approximation.
Inside this space-time, curved 
by the interaction, according to the general covariance,
 quantum wave equations will be proposed.
Then  some simple 
applications will be made, in order to verify the predictions of the theory.

This paper will show the following contents: In Sect. II the operators in the
Schwarzschild metric
will be calculated, in Sect. III, a brief review of the dynamics
will be made, in Sects. IV and V the quantum wave equations will be proposed.
In Sect. VI we will apply the theory to the hydrogen atom, calculating its
energy spectrum and in Sect. VII, the strong interactions will be
considered. In Sect. VIII, the conclusions will be presented.

\section{The metric}

In this section we will calculate the operators ($E$, $\vec p$, $p^2$) needed 
in 
order to write the  wave equations, using the general relativity principles.
As a first step,  a system with spherical symmetry will be considered,
 but the 
basic ideas can be generalized to systems with arbitrary metrics.

We will consider a particle inside a field, that may be described by a 
potential function $V$. The source of the field (a mass for a gravitational 
field or a charge for an electromagnetic field) will have some distribution, described
by a tensor $T_{\m\n}\not= 0$,  in a certain space region.  

Outside of the source distribution, on the empty space (where  
$T_{\m\n}\not =0$)
 if we consider a
 system that  presents spherical symmetry, with a central 
potential $V(r)$, the space-time may be described 
by a Schwarzschild-like metric \cite{lan},\cite{wein},\cite{mould}, 
\beq
ds^2=\xi\ d\tau^2 - r^2(d\theta^2+ \sin^2 \theta\ d\phi^2) - \xi^{-1}dr^2  \ 
 \  ,
\label{I.1}
\eeq 

\ni
where $\xi(r)$ is determined by the interaction potential $V(r)$, and is a 
function only of 
$r$, for a time independent interaction. $\xi(r)$  will be studied in detail 
in Sec. III.  
As we can see in (\ref{I.1}), 
the metric tensor $g_{\m\n}$ is diagonal 
\beq
g_{\m\n}=\pmatrix{\xi&0&0&0\cr 0&-\xi^{-1}&0&0\cr  0&0&-r^2&0\cr
0&0&0&-r^2\sin^2\theta\cr}  \  \  ,
\label{gmn}
\eeq

\ni
and can be written in the form
\beq
g^{\m\n}=h_\m^{-2}\eta^{\m\n}  \  \  ,
\eeq

\ni
where
\beq
\eta_{\m\n}=\pmatrix{1&0&0&0\cr 0&-1&0&0\cr  0&0&-1&0\cr
0&0&0&-1\cr}  \  \  .
\label{hmn}
\eeq

\ni
Using these definitions, we can calculate the operators
\bea
&&\nabla_i=h_i^{-1}{\partial\over \partial x^i }  \  \  ,  \\
&&\vec p=-i\hbar\vec\nabla  \  \  .
\eea

\ni
According to the above expressions, the momentum operator may be defined as
\beq
\vec p= -i\hbar \lb \hat r \sqrt{\xi} {\partial\over\partial r} + 
{\hat\theta\over r}  {\partial\over\partial \theta} + 
{\hat \phi\over r\ {\rm sin} \ \theta} {\partial\over\partial \phi} \rb  \ \ 
,
\label{p}
\eeq

\ni
that in a region with $V=0$ ($\xi=1$) is the usual momentum operator in 
spherical coordinates
\beq
\vec p= -i\hbar \lb \hat r {\partial\over\partial r} + 
{\hat\theta\over r}  {\partial\over\partial \theta} + 
{\hat \phi\over r\ {\rm sin} \ \theta} {\partial\over\partial \phi} \rb 
 \  \ .
\eeq

The energy operator is defined as 
\beq
E = i\hbar\nabla_0 = {i\hbar\over h_0}{\partial\over\partial t}=
{i\hbar\over \sqrt{\xi}}{\partial\over\partial t}
\label{eop}
\eeq

\ni
that for  $V=0$,
\beq
E  = {i\hbar}{\partial\over\partial t}.
\eeq

\ni
The Laplacian is calculated using \cite{wein}
\beq
\nabla^2=(h_1h_2h_3)^{-1}\lb  
{\partial\over\partial x_1} {h_2h_3\over h_1} {\partial\over\partial x_1}+
{\partial\over\partial x_2} {h_1h_3\over h_2} {\partial\over\partial x_2} +
{\partial\over\partial x_3} {h_1h_2\over h_3} {\partial\over\partial x_3}
  \rb  \  \  ,
\label{nab2}
\eeq

\ni
where $h_i$ are given in  (\ref{gmn}), so
\beq
|\vec p|^2=-\hbar^2 \lb {\sqrt{\xi}\over r^2}
{\partial\over\partial r}\lp r^2\sqrt{\xi}{\partial\over\partial r} \rp 
+
{1\over r^2 {\rm sin}\ \theta} 
{\partial\over\partial \theta}\lp {\rm sin}\ \theta 
 {\partial\over\partial \theta}\rp 
+
 {1\over r^2 {\rm sin}^2\theta} {\partial^2\over\partial \phi^2} \rb  \  \  .
\label{p2}
\eeq

If one defines the momentum components as $p_i=-i\hbar\nabla_i$, one
can observe that they are not good operators, these operators
does not commute and are not even  Hermitians, so the 
 definition \cite{pauli}
\beq
p_i={1\over \sqrt{D}} {\partial\over \partial x^i}\sqrt{D}  \  \  ,
\eeq

\ni
where $D=\sqrt{-g}$, with $g={\rm det}({g_{\m\n})}$, will be used. With this 
definition, 
\bea
p_r&=&-i\hbar\lp {\partial\over \partial r} + {1\over r} \rp      \\
p_\theta&=&-i\hbar \lp {\partial\over \partial \theta} + {\rm{cotg}}\theta 
\rp    \\
p_\phi&=&-i\hbar  \lp {\partial\over \partial \phi} \rp  \  \  ,      
\eea

\ni
and the commutation rules are
\bea
&& \lb p_i,q^j  \rb = -i\hbar \delta_i^j  \\ 
&& \lb p_i,p_j\rb = \lb q^i,q^j\rb =0  \  \  .
\eea 

\ni
The Laplacian can be expressed as
\bea
\nabla^2 = && \lp p_r-{i\hbar\over\xi }{\partial\xi \over\partial r} 
\rp
\xi \lp p_r-{i\hbar\over\xi }{\partial \xi\over \partial r} \rp + {1\over r^2}
p_\theta^2 + {1\over r^2\sin^2 \theta}p_\phi^2 \nn \\
&& +{3\over 4} {\partial \xi\over \partial r}- 
{1\over 4} {\partial^2 \xi\over \partial r^2}  \  \  ,
\eea

\ni
that for weak potentials is just
\beq
\nabla^2 =  p_r^2 + {1\over r^2}
p_\theta^2 + {1\over r^2\sin^2 \theta}p_\phi^2  \  \  .
\eeq

With the operators calculated in this section, one can obtain the relativistic
quantum
wave equations. If another symmetry is important (as axial symmetry, for example
), the operators can be obtained in a similar way in the given metric.

\section{Schwarzschild  Dynamics}

In order to obtain the wave equations, two expressions are needed: the energy 
and $\xi(r)$.
 Then, also with the function of setting the notation,
it is useful to make a brief review of the dynamics, 
 in the Schwarzschild metric, and to show how the quantities 
of interest can be expressed.

From  (\ref{I.1}), the proper time is
\beq
d\tau_0=\sqrt{ds^2}=d\tau\sqrt{\xi-{\beta_r^2\over \xi}+r^2\beta_t^2}=
d\tau /\gamma_s  \  \  ,
\eeq

\ni
with
\beq
\gamma_s={1\over\sqrt{\xi-{\beta_r^2\over \xi}+r^2\beta_t^2}}  \  \  ,
\eeq

\ni
where $\beta_r$ and $\beta_t$ are the  the radial and transverse parts  of 
$\vec \beta$, respectively. They are defined as
\bea
&&\vec\beta ={d\vec x\over d\tau}    \  \   ,  \\
&&\beta_r ={d r\over d\tau}    \  \   ,  \\
&&\beta_t = \lb \lp{d\theta\over d\tau}\rp^2+ 
\lp{d\phi\over d\tau}\rp^2\sin^2\theta
\rb^{1/2}
\eea

The principle of least action states that
\beq
S=\int L\ dt=-m_0c^2\int ds=-m_0c^2
\int d\tau \sqrt{\xi-{\beta_r^2\over \xi}+r^2\beta_t^2}   \  \  ,
\eeq

\ni
where $m_0$ is the rest mass of the particle. Then, the Lagrangian can be
expressed as 
\beq
L=-m_0c^2\sqrt{\xi-{\beta_r\over \xi}+r^2\beta_t^2}=-m_0c^2/\gamma_s  \  \  .
\eeq

The momentum four-vector is defined as
\bea
&&p^\m=E_0\beta^\m=E_0\gamma_s(1,\vec\beta)=(p^0,\vec p)   \\
&&p_\m=(\xi \ p^0,-\xi^{-1}\ p^1,- r^2p^2,-r^2\sin^2\theta \  p^3)  \  \  ,
\eea

\ni
where
\beq
E_0=m_0c^2   \   \   .
\eeq

The equivalence principle provides the relation 
\beq
{d\beta^0\over d\tau}=-\Gamma_{\m\n}^0\beta^\m\beta^\n \ \  ,
\eeq

\ni
that gives 
\beq
\gamma_s{dp_0\over d\tau}=E_0\Gamma_{0\n}^\sigma\beta_\sigma\beta^\n=
E_0\lb \Gamma_{00}^1\beta_1\beta^0+\Gamma_{01}^0\beta_0\beta^1  \rb=0  \  \  .
\eeq

\ni
So, the energy defined as
\beq
p_0=E=\xi\gamma_s E_0={m_0c^2\xi\over 
\sqrt{\xi-{\beta_r^2\over \xi}+r^2\beta_t^2}}  \  \  ,
\eeq

\ni
is a constant of motion. The other constant is $L_z=p_3/c$.

In the rest frame of the particle
\beq
{p_0}_\m p_0^\m= -E_0^2=-m_0^2\ c^4  \  \  ,
\eeq

\ni
that is a Lorentz invariant, so
\beq
p_\m p^\m= p^2 c^2-{E^2\over \xi}=-m_0^2\ c^4
\eeq

\ni
and then, the energy relation is
\beq
{E^2\over \xi}=p^2 c^2 + m_0^2 c^4  \  \  ,
\label{en2}
\eeq

\ni
or
\beq
{E\over \sqrt{\xi}}=\sqrt{p^2 c^2 + m_0^2 c^4}  \  \ . 
\label{en}
\eeq

\ni
The expressions (\ref{en2}) and (\ref{en}) will be used to construct the 
Hamiltonian operators.

The term $\xi$ is a function  of $r$, and can be determined if we 
observe (\ref{en})  
\beq
E(\beta=0)=E_0\xi^{1/2}=E_0+V  \  \  ,
\eeq

\ni
that means that in the rest frame of the particle, the energy is due to 
the sum of its 
rest mass with  the potential.
Then
\beq
\xi^{1/2}=1+{V\over E_0}=1+{V\over m_0c^2}  \  \ .
\label{xr}
\eeq

\ni
Comparing with the standard definition of the Schwarzschild mass
\beq
\xi=1-{2\ m_s\over r}=1+{2V\over m_0c^2}+{V^2\over m_0^2c^4} \  \  ,
\eeq

\ni
it is possible to make the identification
\beq
m_s=-{r\over 2}\lp {2V\over m_0c^2}+{V^2\over m_0^2c^4}   \rp  \  \  ,
\eeq

\ni
that for general potentials may be a function of $r$.

For weak potentials,
\beq
\xi\sim 1+{2V\over m_0c^2}  \  \  .
\eeq

\ni
and
\beq
m_s\sim- {Vr\over m_0c^2}   
\eeq

\ni
that are the usual expressions of general relativity, 

\beq
\xi_{G}=\lp 1-{GM\over r\ c^2} \rp^2  \sim 1-{2GM\over r\ c^2} \  \  .
\eeq

\section{Spin-0 particles wave equation}

With the knowledge of the energy  (\ref{en2}) and $\xi(r)$ (\ref{xr}), it 
is possible to formulate   wave equations in the given metric.
The simplest case is to obtain the equation  for spin-0 particles.   
For this purpose, the procedure to be followed is the same one that is
used to determine the Klein-Gordon equation, that is
based in
an operator for $E^2$. Using the relation  (\ref{eop}) 
\beq
 {E^2\over \xi}=-{\hbar^2\over\xi}{\partial^2\over \partial t^2}
\eeq

\ni
 and (\ref{en2})  the quantum wave equation, based on general 
relativity, for spin-0 particles is
\beq
-{\hbar^2\over\xi^2}{\partial^2\Psi\over \partial t^2}=
-\hbar^2c^2\nabla^2\Psi+
 m_0^2 c^4\Psi  \  \  ,
\label{eq0}
\eeq

\ni
with $\nabla^2$ defined in (\ref{nab2}).

This equation can be separated in the standard way, yielding
\beq
\Psi (r,\theta, \phi, t)=u(r)Y_l^m(\theta,\phi)e^{-iEt/\hbar}  \  \  ,
\eeq

\ni
where $Y_l^m(\theta,\phi)$ are the spherical harmonics.
The radial equation is then
\beq
 {\sqrt{\xi}\over r^2}
{\partial\over\partial r}\lp r^2\sqrt{\xi}{\partial\over\partial r} \rp u
+ \lp {E^2\over \hbar^2c^2\xi^2} -  {m_0^2c^2\over \hbar^2} - 
{l(l+1)\over r^2}\rp \ u =0  \  \  ,
\eeq

\ni
and can be solved for a given interaction potential $V(r)$, that
determines $\xi (r)$.

\section{Spin-1/2 particles wave equations}

The next step, is to obtain the analog of the Dirac equation, for spin-1/2 
particles. It can be made in the same way that it was made by Dirac 
\cite{dir1}, using an Hamiltonian with the $\vec \alpha$ and $\beta$ Dirac 
matrices, instead of an energy operator with an square root  
(\ref{en}). Then we have
\beq
{i\hbar \over \xi}{\partial\over\partial t}\Psi=\lp -i\hbar c\ 
\vec \alpha.\vec\nabla +\beta m_0c^2  \rp\Psi \  \  ,
\eeq

\ni
and if we square the operators in both sides of the equation, we obtain 
the wave equation
(\ref{eq0}), that proofs that the procedure used by Dirac, is also valid 
in this case. 

Separating the time dependent part
\beq
T(t)=A\ e^{-iEt/\hbar}  \  \  ,
\eeq

\ni
we will have the spatial equation,
\beq
\lp -i\hbar c\ \vec \alpha.\vec\nabla +\beta m_0 c^2 -{E\over \sqrt{\xi}} 
\rp \psi(\vec r)
=0  \  \  .
\eeq

\ni
Observing the relation 
\bea
\vec\alpha. \vec \nabla &=&{\xi \over r} \vec\alpha. \vec r  
\lb \xi^{1/2}r{\partial \over \partial r} +  \alpha^r
\lp {\alpha^\theta\over r} {\partial \over \partial \theta}+ 
{\alpha^\phi\over r\ {\rm sin}\ \theta} {\partial \over \partial \phi} \rp \rb
\nn \  \  . \\
&& 
\label{eqdd}
\eea

\ni
and the angular part of the operator of  (\ref{eqdd}), 
one concludes   that
the angular part of $\psi(\vec r)$
can be described in terms of the two component spinors 
$\chi_k^\m$, \cite{ros},
\beq
\chi_k^\m=\sum_{m=\pm1/2}C(l,1/2,j;\m-m,m)Y_l^{\m-m}(\theta,\phi)\chi^m
\  \   ,
\eeq

\ni
where $C(l,1/2,j;\m-m,m)$ is a Clebsh-Gordan coefficient, $\chi^m$, a
Pauli spinor and
\bea
&& k=l \ \  \  \  \   \  \   \  \  {\rm for} \  \  j=l-1/2 \  \  , \nn \\
&& k=-l-1  \  \  {\rm for} \  \  j=l+1/2  \  \  ,
\eea

\ni
that gives
\beq
k=\pm (j+1/2)  \  \  .
\eeq

Then, the wave function is expected to have the structure
\beq
\psi=\pmatrix{F(r)\chi_k^\m\cr iG(r)\chi_{-k}^\m}  \  \  ,
\eeq

\ni
with the  $F$ and $G$ functions obeying

\bea
&& \sqrt{\xi} {dF\over dr}+(1+k){F\over r}= 
\lp{E\over \sqrt \xi} +m_0 \rp G  \nn \\  
&& \sqrt{\xi} {dG\over dr}+(1-k){G\over r}= 
-\lp{E\over \sqrt \xi} -m_0 \rp F  \  \  ,
\label{dir}
\eea

\ni
that are equations very similar to the ones obtained from the Dirac theory.
In the following sections, some physical implications of this theory
 will be studied.

\section{The Hydrogen atom}

In this section we will study the behavior of the theory, in a very 
well known system, 
 the hydrogen atom.  In the hydrogen atom, the electron is 
submitted to an electric central potential (obviously with $Z=1$)
\beq
V(r)=-{\alpha Z\over r}  \  \  ,
\label{vcoul}
\eeq

\ni
then the $\xi$ function becomes (where $ep$ means electron-proton)
\beq
\xi_{ep}={\lp 1-{\alpha Z\over m_ec^2 \ r } \rp}^2    \  \  ,
\label{xep}
\eeq

\ni
where $m_e$ is the electron mass. This function represents a space-time
curved by the $ep$ interaction, or how space-time is seen by the 
electron.
At this point we can see an interesting feature of the theory: when general
relativity is used to study the gravitation, the radius where the metric breaks
($\xi=0$) , that is the Schwarzchild radius, $r_s$,
is always negligible, as for example $r=$2.95 Km for the sun. But in the
hydrogen atom (with the parameters of \cite{pdg})
\beq
r_s={\alpha \over m_ec^2  } = 2.818 \ {\rm fm}  \  \  ,
\eeq

\ni
that is the classical radius of the electron, which is obviously not 
negligible.
The estimated radius of the proton is about 0.9-1.0 fm, so, it is located 
inside the horizon of events. Then, the 
electric charge will be confined inside this region by a trapping surface, as 
defined in \cite{pen}, and outside, only effects of the total 
charge can be probed by the electron, and no information about the 
inner structure can be obtained.

Inserting $\xi_{ep}$ from the expression (\ref{xep}) in the equations 
(\ref{dir}) we obtain the equations, but valid only for $r>r_s$. Inside the 
horizon of events,
 the metric is not the same, the energy-momentum tensor $T_{\m\n}$ 
determined by the charge and matter distributions
must
be considered. In this paper
we shall study only the outside behavior. It is illustrative to study the 
approximation 
\bea
  {dF\over dr}+(1+k){F\over r}&=& 
\lp {E\over 1+V/E_0} +m_e \rp G \nn \\
&\sim & 
\lp E-{EV\over E_0} +m_e \rp G  \  \  , \nn  \\  
  {dG\over dr}+(1-k){G\over r}&=& 
-\lp{E\over 1+V/E_0} -m_e \rp F \nn \\
& \sim &
-\lp  E-{EV\over E_0} -m_e \rp F  
\label{dir2}
 \  \  ,
\eea

\ni
where we neglected the $V/E_0$ and higher 
order terms. As we can see, in this theory, 
the Dirac theory is recovered only for $E/E_0\sim 1$ (lower moments),
\bea
&&  {dF\over dr}+(1+k){F\over r}= 
\lp E-V +m_e \rp G  \nn \\  
&&  {dG\over dr}+(1-k){G\over r}= 
-\lp E-V -m_e \rp F  \  \  .
\eea

\ni
We must remark that in the
approximation of  equation (\ref{dir2}), the metric divergence at  $r=r_s$ 
(that is not a physical 
 singularity, it appears from the choice of the coordinate 
system)
is removed, then $\xi\sim$1 and $r_s$ does not exist.
So, in this case,  the wave functions must be controlled only at the physical
singularity, at the
origin. Considering  the Frobenius method, the solutions 
are of the type 
\bea
&& F=\rho^s\sum_{n=0}^N a_n \rho^ne^{-\rho}   \  , \nn  \\
&& G=\rho^s\sum_{n=0}^N b_n \rho^ne^{-\rho}  \  . 
\label{frob}
\eea

\ni
where $\rho=\beta r$, and after some manipulations 
one finds 
\beq
\beta=\sqrt{m^2c^4-E^2}
\eeq

\ni
 and
\beq
s=\sqrt{k^2-{\gamma^2E^2\over m^2}}  \  \  .
\eeq

\ni
Observing the solutions (\ref{frob}) one can see that
 $F(r\sim 0)$ and $G(r\sim 0)$ are sums of terms of the type
$Ce^{-\rho}\rho^{s+m} (r\sim 0)$, and then, $\psi(r=0)$=0.
So, the effect of the approximation, is to remove the horizon of events, and 
to extend the solution to the region $r<r_s$. 
Considering the exact solution of the equation, this behavior is valid only for
$r>r_s$, and an horizon of events exists at this surface with the
proprieties described above. But one must remark that
at the atomic level, an approximation of the order of 2.8 fm is
not so bad, in this region the wave functions are almost negligible
and for practical purposes this approximation is reasonable.

From these expressions, it is possible to  calculate the energy spectrum.
Using the standard methods \cite{ros}, \cite{scf}, \cite{bj} one obtains
\beq
\lp m_e^2c^4-E^2   \rp a^2={E^4\gamma^2\over m_e^2c^4}  \  \  ,
\eeq

\ni
that has the solutions
\beq
E^2=m_e^2c^4\lb {-1 \pm \sqrt{1+4\gamma^2/ a^2} \over 2\ \gamma^2} \rb  \ \
,
\label{rrt}
\eeq

\ni
with
\beq
\gamma= {\alpha\over\hbar c}  \  \  ,
\eeq

\ni
and
\beq
a=a(n)=n-\lp j+1/2\rp +\sqrt{\lp j+1/2\rp^2-\gamma^2}  \  \  .
\eeq

\ni
Using only the positive root solutions of (\ref{rrt})
  the hydrogen atom spectrum is

\beq
E_n=
 m_ec^2\sqrt{2\over {1+  \sqrt{1+4\gamma^2/a^2} }}  \  \  .
\label{ehr}
\eeq

\ni
Adopting the expansion ($\gamma^2/a^2$ is small)
\beq
\sqrt{1+4{\gamma^2\over a^2}}\sim 1+2{\gamma^2\over a^2}-2{\gamma^4\over a^4}+
4{\gamma^6\over a^6}-10{\gamma^8\over a^8}+28{\gamma^{10}\over a^{10}}+....
\eeq

\ni
the energy spectrum (\ref{ehr}) can be rewritten as
\beq
E_n={m_ec^2\over \sqrt{1+\gamma^2/a^2-\gamma^4/a^4+2\gamma^6/a^6-
5\gamma^8/a^8+...}}  \  \  ,
\label{bsp}
\eeq

\ni
where we can find explicitly
the corrections of the energy levels,
 due to general relativistic effects,
if compared with the standard \cite{scf}, \cite{bj} relativistic spectrum
\beq
E_n={m_ec^2\over \sqrt{1+\gamma^2/a^2}}  \  \  ,
\label{hsp}
\eeq
 
\ni
that can be obtained from the Dirac equation or from the Sommerfeld model
\cite{som}.

Considering now the spectrum obtained from the exact solution of  
(\ref{dir}), without the approximations made in  (\ref{dir2})
\beq
E_N= m_ec^2 \sqrt{{1\over 2}- {N^2\over 8\alpha^2} + 
{N\over 4\alpha}\sqrt{{N^2\over 4\alpha^2}+2}}   \  ,
\label{sbr}
\eeq

\ni
we will compare the theoretical results with the experimental data and with 
the ones obtained with the Dirac theory.

Some numerical values are shown in Table I, 
the experimental results \cite{nist} for the 
differences between the energies $E(n,l,j)$
and the ground-state energies $E_1$, for the
hydrogen atom
and  deuterium,  
the corresponding values calculated with the Dirac theory (\ref{hsp}),
and the results calculated in this work, with (\ref{sbr}).
Observing the table, one can see that the accord of both theories
with the hydrogen experimental data is very good, but the results from
 (\ref{sbr}) are closer to the experimental data then the 
results from  (\ref{hsp}).
One must remark
that spherical symmetry is not exact in the hydrogen atom as the proton 
mass is finite, but  
 with a heavier nuclei, this symmetry
is a better approximation, so it is interesting to observe the 
 deuterium data.
Comparing the results form the Dirac theory  (\ref{hsp}), one can see 
a better accord, and the deviations from the data are of the order 
of 0.027\%. Considering the spectrum  (\ref{sbr}), the 
 deviations are of the order
of 0.005\%,
 approximately five times smaller.

\begin{table}[hbt]
\begin{center}
\caption{Experimental energy levels (eV) for the hydrogen atom, for the 
deuterium \cite{nist}, and the theoretical ones, calculated with the Dirac theory 
(\ref{hsp}) and with  (\ref{sbr}).} 
\begin{tabular} {|c|c|c|c|c|}
\hline

		& Hydrogen 	& Deuterium 	
& Dirac 	& Eq. (\ref{sbr}) 
		\\ \hline
$E_1$	& -13.59844 & -13.60214 & -13.60587 & -13.60298 
		\\  \hline
$ E(2,0,1/2)-E_1$& 10.19881  & 10.20159  & 10.20444  & 10.20172	
	\\ \hline 
$ E(3,0,1/2)-E_1$& 12.08750  & 12.09079  & 12.09413  & 12.09127
	\\ \hline
$ E(4,0,1/2)-E_1$& 12.74854  & 12.75201  & 12.75551  & 12.75263	
	\\ \hline
$ E(5,0,1/2)-E_1$& 	13.05450  & 13.05806  & 13.06164  & 13.05875
	\\ \hline
\end{tabular}
\end{center}
\end{table}

\section{Strong interactions}

In this section,  the implications of the
theory, when strong interactions are taken into account
will be studied. 
The simplest system possible is the $NN$ interaction. If, as a first
approximation, only the long range part of the potential would be considered,
it should be dominated by
the one pion exchange contribution  (Yukawa potential), 
\beq
V(r)=g^2{e^{-\mu r}\over r}
\eeq

\ni
where $g^2=13.40$  is the $NN$ coupling constant and $\m$ is the 
pion mass.
As $V(r)$ is a function only of $r$, if we locate one nucleon at the origin,
we would have 
\beq
\xi_{NN}=\lp 1-g^2{e^{-\m r}\over m_Nc^2 \ r } \rp^2  \  \  .
\label{xist}
\eeq

\ni
However, as
it is well known,  the $NN$ potential is not central,
and
there are other contributions, such as the tensor part \cite{par}, \cite{reid},
\cite{ar} that arises from
more complex processes (two pion exchange \cite{man} and others).
 In order to make some estimates,
some symmetrical cases of the Reid \cite{reid} potentials can be used, as
they  are phenomenological ones 
and can give a first idea of the Schwarzschild radius.
The potentials are superpositions of Yukawa type terms,
\bea
&&V(^1S)=-h\lp e^{-x}+39.633e^{-3x}  \rp / x    \  \   ,  \\
&&V(^1D)= -h\lp e^{-x}+4.939e^{-2x}+154.7e^{-6x}  \rp / x      \  \  ,  \\
&&V(^1S)_s=-\lp h e^{-x}+1650.6e^{-4x}-6484.2e^{-7x}\rp  / x    \  \   , \nn 
\\
&& 
\eea

\ni
where $x=0.7 \ r \ {\rm fm}^{-1}$, $r$ is the relative radius and $h$=10.463 
MeV.
 Inserting these 
potentials in  (\ref{xr}) and equating it to 0, we will have
\bea
&&r_s(^1S)=0.33    \  {\rm fm}  \  \  ,  \\
&&r_s(^1D)=0.44    \  {\rm fm}  \  \  ,  \\
&&r_s(^1S)_s\sim 0.33    \  {\rm fm}  \  \  .  
\eea

\ni
Thus,  the part of the source of the strong force that is inside  the horizon 
of events will be submitted to the  the trapping effect \cite{pen}, that 
  prevents the escape of
any matter and radiation, what means 
radial collapse of the source of
the strong forces.

It must be noted that the potentials containing spin dependent terms and
others were not considered, fact that would break the spherical symmetry.
However, in a first approximation, these terms can be considered as 
corrections to the central potential. 
But even if we consider these terms, with another metric, as
for example an axial 
symmetric metric, the trapping surface would still exists, confirming the 
present conclusions, and, only giving a more accurate estimate of the size of 
the confining
region. Collapse is not an exclusive feature of spherical symmetric systems,
as it was stated in \cite{pen}, deviations from spherical symmetry cannot 
prevent space-time singularities from arising.

As we can see, when strong interactions are considered, the horizon of events
is located in a radius that is not negligible. 
The preceding example gives an estimate in the range of 0.3-0.5 fm.
This fact suggests that the 
quark confinement may be understood from these results. To understand the 
mechanism, let us consider that the source of the strong force 
obeys some matter distribution. Each element of this matter distribution,
(that may be a quark, but in general, this assumption is not necessary) 
with mass $m_0$, suffers the action of an attractive strong force. 
Some models \cite{a}, \cite{b},\cite{c}, \cite{d}
 consider central Coulombic potentials of the type (\ref{vcoul})
 to describe the 
effective interaction to which the quarks are submitted inside an hadron.
A good example is the Cornell model \cite{corn}, that
with a linear plus Coulomb
central potential
\beq
V(r)=-{\alpha\over r}+{r\over a^2}  \  \  ,
\label{vcrn}
\eeq

\ni
with the parameters $a\sim$ 2.34 ${\rm GeV}^{-1}$ and $\alpha\sim$ 
0.5, is able to describe the $J/\psi$ and the $\Upsilon$.

Thinking in terms of constituent quarks
it is possible to use the proposed theory  to give a
description of some hadrons. Table II shows an estimate of $\alpha$
(for a Coulomb potential) and of the constituent quark masses ($m$)
in order to obtain the masses ($M$) of the nucleon and the $J/\psi$ and 
$\Upsilon$
mesons. The experimental values of these masses are also shown.
These constants  define the value of $r_s$, inside of which the
quarks are expected to be confined. The values of $r_s=$ 0.83 fm for a nucleon
and $0.05-0.11$ fm for heavy mesons are very reasonable and show  that for
heavier quarks, the values of $r_s$ are smaller. Table II was constructed 
with the objective of
 giving an  idea of the magnitude of the constants, but a detailed
description of the observed hadrons must consider additional terms in the 
potential.

\begin{table}[hbt]
\begin{center}
\caption{Values of the masses $M$ compared with the experimental ones 
\cite{pdg}
for some systems.
The calculations are made considering Coulomb potentials 
with the parameters
 $\alpha$, $m$ and $r_s$.} 
\begin{tabular} {|c|c|c|c|c|c|}
\hline

		& $m$(GeV) 	& $\alpha$ 	
& $r_s$(fm) 	 & $M$(GeV) & $M_{\rm exp}$(GeV)
		\\ \hline
Nucleon($qqq$)& 0.38  & 1.60  & 0.83 & 0.938    & 0.938 (proton)\\ \hline
$J/\psi (c\bar c)$& 1.79  & 1.00  & 0.11 &3.10  & 3.10  	\\ \hline
$ \Upsilon (b\bar b)$& 	5.50  & 1.05  & 0.05 & 9.47  & 9.46 	\\ \hline
\end{tabular}
\end{center}
\end{table}

\ni
With these results, it is possible to calculate \cite{bog}, \cite{clos}
\beq
{g_A\over g_V}= {5\over 3} <\sigma_Z>={5\over 3}(1-2\ \delta)  \  ,
\eeq

\ni
where 
\beq
\delta={{2\over 3}\int |G(r)^2|dr\over
\int\lp |F(r)|^2+|G(r)|^2  \rp dr}=0.059
\eeq

\ni
where a nucleon composed of tree quarks with $j_z=1/2$ is considered.
So, $g_A/g_V$=1.47, what shows a 17\% deviation from the experimental
result that is 1.259. The magnetic moments of the proton and the neutron
may also be calculated
\bea
\m_p&=&{(1-\delta)m_p\over E_0}=2.82  \nn \\
\m_n&=&- {2\over 3} {(1-\delta)m_p\over E_0}        =-1.88  \   ,
\eea

\ni
that are in good agreement with the experimental results $\m_p$=2.79
and $\m_n$=-1.91 \cite{pdg}.

If the whole quark content of the hadron is located inside $r_s$
(now, spherical symmetry is a good choice), 
the classical description of such a system would predict the collapse 
of the whole matter in the singularity located at $r=0$. However, as we are 
dealing with a  quantum system, the uncertainty principle will prevent 
this collapse. Consequently, there are two opposite effects acting on the
elements of matter, and the resulting effect will be radial oscillations.
This effect, in a flat Minkowski space-time
may be described by  effective potentials of the form
\beq
V_{\rm eff}\sim a_0+a_1r+a_2r^2+...  \  \  .
\label{potc}
\eeq

\ni
There is no surprise why some authors \cite{x},
\cite{w},   explain the 
hadronic structure with models
based on harmonic oscillator quark models or with linear potentials of the type
$\lambda r$\cite{zz}, \cite{corn}.
Inside the horizon of events, $V>E_0$, so Eq. (\ref{dir}) can be expanded
as
\bea
&& \lp 1- {\alpha\over m_0 r}  \rp {dF\over dr}+(1+k){F\over r} \nn \\
&& \sim \left\{ -E\lb {m_0r\over\alpha} + \lp {m_0r\over\alpha} \rp^2  
-\lp {m_0r\over\alpha} \rp^3+... \rb +m_0  \right\} G \nn \\
&& \lp 1- {\alpha\over m_0r}  \rp {dG\over dr}+(1-k){G\over r} \nn \\
&&
\sim -\left\{ -E\lb {m_0r\over\alpha} + \lp {m_0r\over\alpha} \rp^2  
-\lp {m_0r\over\alpha} \rp^3+... \rb -m_0  \right\} F \  \  ,
  \nn \\
\label{dirq}
\eea

\ni
where terms similar to the effective potential
(\ref{potc}) appears.
So, considering a Coulombic potential (fact that is not strictly necessary,
other kind of potentials may present similar proprieties)
with the correct parameters, confinement effects may occur, fact that
is widely used, with 
 the addition of confining potentials 
in flat space-times.

\section{Summary and conclusions}

In this paper,  quantum wave equations, based on 
 the general relativity,
in a Schwarzschild-like metric, have been obtained. 
Investigating the hydrogen atom 
spectrum, the approximate expression resulting from the theory (\ref{ehr})
is in accord with the experimental values, and
shows a small improvement due to 
the general relativistic corrections, when compared 
with the standard relativistic spectrum (\ref{hsp}). 
Although, if the exact solution is considered, the
corrections are not so small and
 (\ref{sbr}) gives a 
significant improvement of the accord with the experimental data, specially
with the deuteruim spectrum.

An interesting feature of the theory, is that  in the  Schwarzschild metric,  
the horizon of events  appears for $r=r_s$, with a value that is not 
negligible, 
as it 
happens when the  gravitational interaction is considered. 
 When considering the strong interaction,
$r_s$ shows a region inside the hadron, where confinement 
arises. From this theory, confinement may be considered as an intrinsic 
propriety of the space-time, that when interactions with large coupling 
constants are considered, generates trapping surfaces.
On the other hand,  no collapse for $r=0$ is expected, the uncertainty 
principle
forbids it.
 $\xi$ as defined in  (\ref{xr})
 is a function of $\alpha/ m$, so, in 
Nature, $\alpha$ and $m$ are such that the confinement conditions are filled
and in a region with the observed size (some examples are shown in Table II).

Thinking matter as composed of small black holes may seem a weird idea,
but no one has actually seen a quark, or an element of strongly interacting 
matter, and in this sense, a black hole is quite reasonable.
At astronomical level also, it is possible to imagine systems that cannot be 
seen, due
to the curvature caused by electromagnetic or strong forces, and maybe
giving 
an important contribution to the mass of the universe.

The most important feature of the theory, is the fact that the 
 insertion of general 
relativistic aspects in the quantum theory generates results that 
are in accord with 
the phenomenology of the considered systems.

\begin{acknowledgments}
I wish to thank  professor M. R. Robilotta.
This work was supported by FAPESP.
\end{acknowledgments}




\end{document}